\documentclass[letter]{aa}
\usepackage{rotating}
\usepackage{graphicx}
\usepackage{txfonts}
\begin{document}

\title{HD\,45314: a new $\gamma$\,Cas analog among Oe stars\thanks{Based on observations collected with {\it XMM-Newton}, an ESA Science Mission with instruments and contributions directly funded by ESA Member States and the USA (NASA), and observations collected at the European Southern Observatory (La Silla, Chile) and the Observatoire de Haute Provence (France).}}
\author{G.\ Rauw\inst{1} \and Y.\ Naz\'e\inst{1}\fnmsep\thanks{Research Associate FRS-FNRS (Belgium)} \and M.\ Spano\inst{2} \and T.\ Morel\inst{1} \and A.\ ud-Doula\inst{3}}
\offprints{G.\ Rauw}
\mail{rauw@astro.ulg.ac.be}
\institute{Groupe d'Astrophysique des Hautes Energies, Institut d'Astrophysique et de G\'eophysique, Universit\'e de Li\`ege, All\'ee du 6 Ao\^ut, B\^at B5c, 4000 Li\`ege, Belgium
\and Observatoire de Gen\`eve, Universit\'e de Gen\`eve, 51 Chemin des Maillettes, 1290, Sauverny, Switzerland
\and Penn State Worthington Scranton, Dunmore, PA 18512, USA} 
\date{Received date / Accepted date}
\abstract{Oe stars possibly form an extension to higher temperatures of the Be phenomenon, but it is still unclear whether these stars have disks.}{X-ray spectra could provide hints for interactions of the star with a putative surrounding disk.}{We obtained XMM-Newton observations of two Oe stars, HD\,45314 and HD\,60848. Spectra and light curves were extracted and analysed. Optical spectra were also obtained to support the X-ray observations.}{We find that both stars display very different X-ray properties. Whilst HD\,60848 has an X-ray spectrum and emission level typical for its spectral type, HD\,45314 displays a very hard X-ray emission, dominated by a thermal plasma with $kT \sim 21$\,keV. Furthermore, HD\,45314 displays count rate variations by a factor 2 on timescales of $\sim 10^3$\,s and a high $\log{\frac{ L_{\rm X}}{L_{\rm bol}}} = -6.10 \pm 0.03$.}{The X-ray properties of HD\,45314 indicate that this star is a new member of the class of $\gamma$\,Cas analogs, the first one among the original category of Oe stars.}
\authorrunning{G. Rauw et al.}
\titlerunning{HD\,45314, a new $\gamma$\,Cas analog}
\maketitle
\section{Introduction}
The class of Oe stars was defined by Conti \& Leep (\cite{CL}) as O stars displaying emission lines of the H\,{\sc i} Balmer series as well as of lower ionization elements such as He\,{\sc i} and Fe\,{\sc ii}, but that do not show typical Of emission lines such as He\,{\sc ii}\,$\lambda$\,4686 and N\,{\sc iii}\,$\lambda$\,4634-40. These stars were commonly considered to be massive analogs of Be stars. However, Negueruela et al.\ (\cite{NSB}) revised the spectral types of most Oe stars, arguing that previous classifications were too early because of infilling of the He\,{\sc i} lines. Most Oe stars would thus have true spectral types in the range O9 -- B0 (the earliest member of the Oe category is the O7.5\,IIIe star HD\,155806, Negueruela et al.\ \cite{NSB}). 
As for Be-type stars, the emission lines of Oe stars frequently display a double-peaked morphology and both their line strength and profile (e.g.\ the intensity ratio of the blue and red peaks) undergo strong variations (e.g.\ Rauw et al.\ \cite{ibvs}). By analogy with Be stars (Porter \& Rivinius \cite{PR}), these features are traditionally attributed to a geometrically thin decretion disk-like equatorial wind undergoing density variations. 
However, unlike the situation of Be stars, no direct images of the disks of Oe stars have been obtained so far, and the mere existence of disk-like winds around Oe stars has been questioned by Vink et al.\ (\cite{Vink}). Indeed, the linear spectropolarimetric observations of Vink et al.\ (\cite{Vink}) showed the depolarization effect across the H$\alpha$ emission line, that is expected in the presence of a disk, for only one case out of six\footnote{Two of the Oe stars of Vink et al.\ (\cite{Vink}) had H$\alpha$ in absorption and were thus unlikely to feature a dense circumstellar disk at the time of their observations. See also Naz\'e et al.\ (\cite{NRU2}).}: HD\,45314, which is one of our targets here. 

Because the X-ray emission of single early-type stars arises in their winds (e.g.\ G\"udel \& Naz\'e \cite{GN}), X-ray observations could help shed some light on the circumstellar environment of Oe stars. Previously, two Oe stars have been observed in X-rays, HD\,119682 (O9.7-B0\,Ve\footnote{This star was not classified as one of the original Oe stars.}, Rakowski et al.\ \cite{Rakowski}) and HD\,155806 (O7.5\,IIIe, Naz\'e et al.\ \cite{NRU}). On the one hand, HD\,155806 has a very soft X-ray spectrum with two plasma temperatures of $kT = 0.2$ and $0.6$\,keV, as well as broad emission lines and an emission level consistent with normal O-type stars (Naz\'e et al.\ \cite{NRU}). On the other hand, HD\,119682 (O9.7e) displays a very hard X-ray spectrum with a dominant plasma component at kT $\sim 10$\,keV, making this object a so-called $\gamma$\,Cas analog (Rakowski et al.\ \cite{Rakowski}). 

The definition of $\gamma$\,Cas analogs is based on their remarkable X-ray properties (e.g.\ Smith et al.\ \cite{Smith1}) that we shortly summarize here. Their $\frac{L_{\rm X}}{L_{\rm bol}}$ ratio is a factor 10 higher than for normal OB stars (which have $\frac{L_{\rm X}}{L_{\rm bol}} \sim 10^{-7}$, e.g.\ Bergh\"ofer et al.\ \cite{Berghoefer}, Naz\'e \cite{Naze}), but at least an order of magnitude below that of Be X-ray binaries. Variability is found on many timescales up to several months. The fastest variations, so-called flares, occur on timescales from several seconds to minutes. Their X-ray spectrum consists of 3 - 4 thermal plasma components, but is dominated by the hot plasma (around 14\,keV for $\gamma$\,Cas, but sometimes exceeding 30\,keV for HD\,110432, Smith et al.\ \cite{Smith2}). 

There are two main scenarios for the origin of the X-ray emission of $\gamma$\,Cas analogs: accretion of the Be disk by a white dwarf companion\footnote{$\gamma$\,Cas is indeed a binary system with an unseen companion that might well be a white dwarf, see Smith et al.\ (\cite{Smith1}), although the orbit is wide, making accretion inefficient (Torrej\'on et al.\ \cite{TSN}).}, or magnetic star-disk interactions (see Smith et al.\ \cite{Smith1} and references therein).

To enlarge the sample of X-ray observations of Oe stars, we obtained {\it XMM-Newton} observations of HD\,45314 (PZ\,Gem) and HD\,60848 (BN\,Gem). Both stars were on the list of Oe stars of Conti \& Leep (\cite{CL}). Originally classified as O9?pe and O8\,V?pe, they were reclassified by Negueruela et al.\ (\cite{NSB}) as B0\,IVe and O9.5\,IVe for HD\,45314 and HD\,60848, respectively. 

\section{Observations and data reduction \label{observations}}
HD\,45314 and HD\,60848 were observed with {\it XMM-Newton} (Jansen et al.\ \cite{Jansen}) in April 2012 (ObsIDs 0670080301 and 0670080201 respectively, see Table\,\ref{journal}). The EPIC cameras (Turner et al.\ \cite{MOS}, Str\"uder et al.\ \cite{pn}) were operated in full-frame mode and were used with the thick filter to reject optical and UV photons. The data were processed with the SAS software version 11.0. Both observations were affected by high-background events (so-called soft-proton flares). For HD\,45314, these high-background events took the form of two relatively short flares, whereas for HD\,60848 only the first $\sim 6$\,ks of the integration benefit from a low background level. Except when stated otherwise, our subsequent analysis refers to data products obtained after rejecting the high-background episodes. Both Oe stars are detected, although with very different count rates (Table\,\ref{journal}). We extracted the spectra and light-curves in three energy bands, medium (1.0 -- 2.0\,keV), hard (2.0 -- 8.0\,keV) and total (0.5 -- 10\,keV), for each star and each of the three EPIC instruments. 
For the brighter of our targets, HD\,45314, we also processed the data collected with the RGS reflection spectrometer (den Herder et al.\ \cite{RGS}). However, given the relatively modest count rate, the short integration time and the properties of HD\,45314's X-ray spectrum (see below), the RGS spectra do not provide useful data. 
\begin{table}
\caption{Journal of the observations.\label{journal}}
\begin{tabular}{c c c c c c}
\hline
\multicolumn{6}{c}{\it XMM-Newton}\\
\hline
Target & Date & Exposure & Clean & MOS & pn\\
       & JD-2450000 & (ks) & (ks) & (ct\,s$^{-1}$) & (ct\,s$^{-1}$) \\
\hline 
HD\,45314 & 6031.94 & 24.3 & 20.0 & 0.136 & 0.198 \\
HD\,60848 & 6020.21 & 30.2 & 6.0  & 0.007 & 0.033 \\
\hline
\end{tabular}
\tablefoot{Columns 3 and 4 provide the actual exposure time of the EPIC-pn camera and the remaining time after discarding high-background episodes. Columns 5 and 6 provide the sum of the MOS1 and MOS2 count rates and the pn count rates.}
\begin{tabular}{c c c c c}
\hline
\multicolumn{5}{c}{Optical spectroscopy}\\
\hline
Target & Date & Exp. time & Instrument & EW(H$\alpha$)\\
       & JD-2450000 & (min) &              &  (\AA)       \\
\hline 
HD\,45314 & 5995.29 & 10 & SOPHIE  & $-22.7$ \\
HD\,45314 & 5997.51 & 18 & CORALIE & $-23.2$ \\
HD\,45314 & 6032.48 & 22 & CORALIE & $-22.3$ \\
HD\,60848 & 6019.50 & 25 & CORALIE & $-4.5$ \\
\hline
\end{tabular}
\tablefoot{In Col.\ 5, negative equivalent widths indicate emission.}
\end{table}

In support of the X-ray observations, we obtained several optical echelle spectra with the CORALIE and SOPHIE spectrographs at the Swiss 1.2\,m Leonhard Euler Telescope at La Silla and the 1.93\,m telescope at Observatoire de Haute Provence (Table\,\ref{journal}). CORALIE is an evolution of the ELODIE spectrograph (Baranne et al.\ \cite{Baranne}) and covers the spectral range from 3850 to 6890\,\AA. The SOPHIE spectrograph (Perruchot et al.\ \cite{Sophie}) covers the domain from 3870 to 6940\,\AA. The data were first reduced with the pipelines of the spectrographs and were subsequently normalized by fitting a spline function through a large number of carefully selected continuum windows. Equivalent widths (EWs) of the H$\alpha$ emission were measured and both stars are found to be at the lower end of the range of values observed by Rauw et al.\ (\cite{ibvs}), which span between $-20$ and $-39$\,\AA\ for HD\,45314, and between $-6$ and $-15$\,\AA\ for HD\,60848.

\begin{figure*}[h!tb]
\begin{minipage}{8cm}
\begin{center}
\resizebox{8cm}{!}{\includegraphics{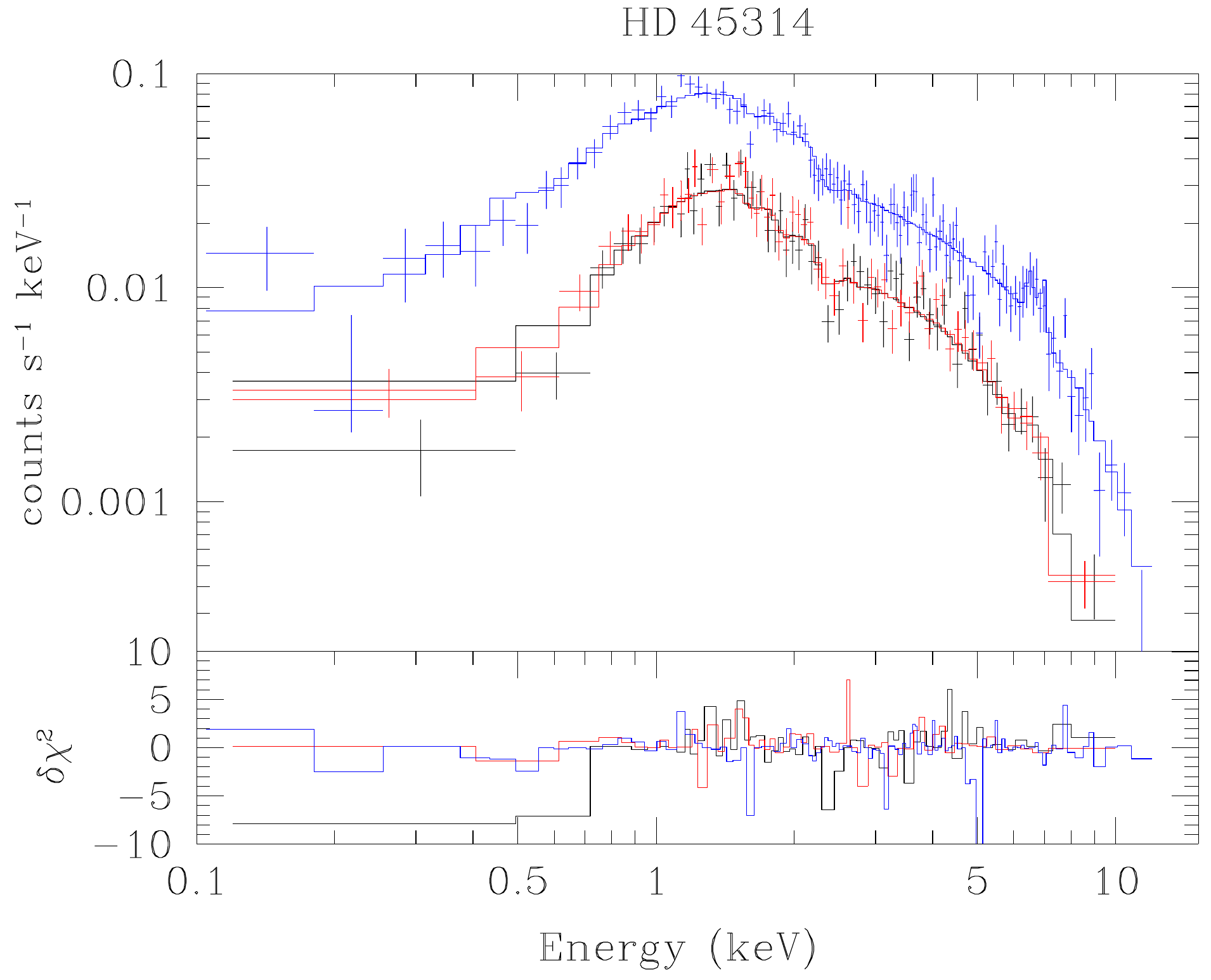}}
\end{center}
\end{minipage}
\hfill
\begin{minipage}{8cm}
\begin{center}
\resizebox{8cm}{!}{\includegraphics{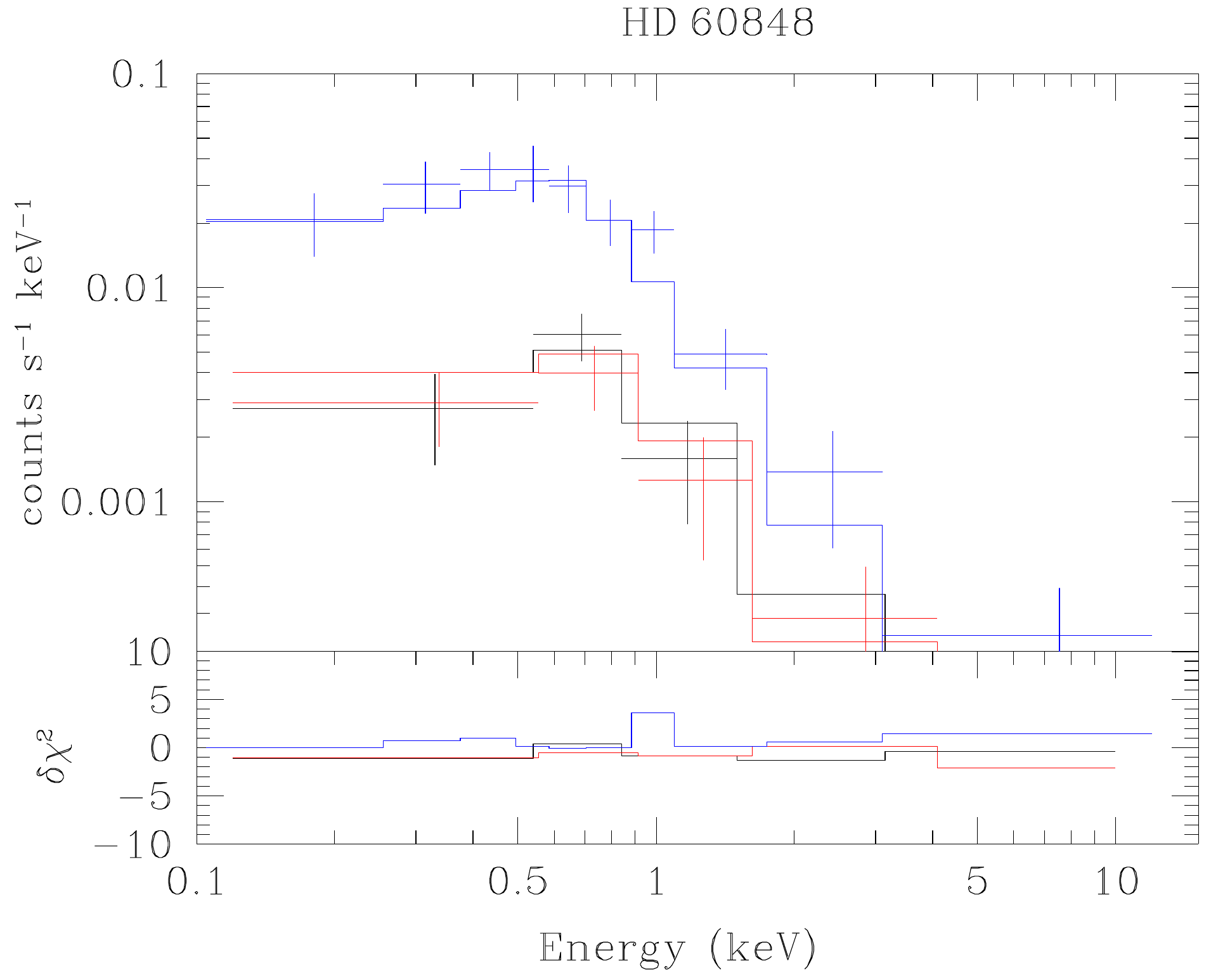}}
\end{center}
\end{minipage}
\caption{EPIC spectra (along with the best-fit model) of the two Oe stars studied here. The black, red and blue lines refer to the MOS1, MOS2 and pn spectra, respectively. The horizontal and vertical scales are the same in both panels. For HD\,45314, the model includes a fluorescent Fe K line.\label{fig1}}
\end{figure*}

\section{X-ray spectra}
As can be seen in Fig\,\ref{fig1} and from the count rates in Table\,\ref{journal}, HD\,45314 (left panel) is a bright and hard source, whereas HD\,60848 (right panel) is much weaker and displays a much softer spectrum. For HD\,45314, 58\% of the received X-ray photons in the 0.5 - 10\,keV energy range, have energies above 2\,keV, whilst this fraction amounts to only 1\% for HD\,60848.

To further quantify these differences, we analysed the EPIC spectra of our targets with the {\tt xspec} software (version 12.7, Arnaud \cite{Arnaud}). For this purpose, we used optically thin thermal plasma ({\tt apec}, Smith \& Brickhouse \cite{apec}) models with solar abundances according to Anders \& Grevesse (\cite{AG}), as well as non-thermal power-law models. The interstellar neutral-hydrogen column density was taken from the compilation of Gudennavar et al.\ (\cite{ISM}). To account for a possible additional absorption of the X-rays by the stellar winds, we also included an ionized-wind absorption model (Naz\'e et al.\ \cite{wind}), when this was necessary to achieve a good fit.  

For HD\,45314, the best fit is obtained with a model of the kind {\tt phabs*wind*apec}, i.e.\ containing a single-temperature plasma component (see Table\,\ref{fitspec}). The most remarkable result concerns the best-fit temperature, $kT = 21$\,keV, which is much higher than in other O-type stars, which typically have $kT \simeq 0.6$\,keV (e.g.\ Naz\'e \cite{Naze}). There is no need for a second temperature component. An almost equally good fit was achieved for an absorbed power-law model ($\chi^2_{\nu} = 1.20$, $\Gamma = 1.4$). For both types of models, the observed and ISM absorption-corrected fluxes amount to $1.21 \times 10^{-12}$ and $1.33 \times 10^{-12}$\,erg\,cm$^{-2}$\,s$^{-1}$, respectively. The EPIC-pn spectrum reveals a weak Fe K line redwards of the Fe {\sc xxvi} Ly$\alpha$ line. Including a Gaussian to fit this line slightly improves the fit ($\chi^2_{\nu} = 1.15$) and yields a best-fit energy of $6.48^{+.14}_{-.23}$\,keV, consistent with fluorescence from highly ionized Fe and an equivalent width of 0.11\,keV (33\,m\AA). All these properties, except for the fact that a single plasma component is sufficient to fit the spectra, indicate that HD\,45314 is a new $\gamma$\,Cas analog, the first one among the stars of the original Oe class of Conti \& Leep (\cite{CL}). 

The situation is quite different for HD\,60848. Although the spectra of the latter source are of poor quality, a good fit requires a two-temperature plasma, but both plasma temperatures are below 1\,keV. A slightly better fit is obtained if one of the plasma components is replaced by a steep power-law component ($\chi^2_{\nu} = 1.02$, $kT = 0.23$, $\Gamma = 2.9$). The observed and absorption-corrected fluxes are about a factor 35 lower than for HD\,45314. The X-ray spectrum of HD\,60848 is quite typical for normal O-type stars (e.g.\ Naz\'e \cite{Naze}). 

\begin{table*}[h!tb]
\caption{Best fits of the EPIC spectra of the two Oe stars with thermal plasma models.\label{fitspec}}
\begin{tabular}{c c c c c c c c c c}
\hline
Star & d.o.f.& $\chi^2_{\nu}$ & N$_{\rm H}$  & $\log{N_{\rm wind}}$ & $kT_1$ & $kT_2$ & norm$_2$/norm$_1$ & $f_X^{\rm obs}$            & $f_X^{\rm un}$ \\
     &       &          & ($10^{22}$\,cm$^{-2}$) & (cm$^{-2}$) & (keV) & (keV) & & (erg\,cm$^{-2}$\,s$^{-1}$) & (erg\,cm$^{-2}$\,s$^{-1}$) \\
\hline
HD\,45314 & 225 & 1.16 & 0.19 (fixed) & $21.0^{+.2}_{-.2}$ & $21^{+8}_{-5}$ & -- & -- & $(1.21^{+.02}_{-.02}) \times 10^{-12}$ & $(1.33^{+.04}_{-.04}) \times 10^{-12}$ \\
\vspace*{-3mm}\\
HD\,60848 & 16 & 1.21 & 0.03 (fixed) & -- & $0.12^{+.08}_{-.03}$ & $0.83^{+.51}_{-.15}$ & $0.13 \pm 0.11$ & $(3.2^{+.3}_{-.3}) \times 10^{-14}$ & $(3.8^{+.5}_{-.5}) \times 10^{-14}$ \\
\vspace*{-3mm}\\
\hline
\end{tabular}
\tablefoot{The last two columns yield the observed and absorption-corrected fluxes over the 0.5 -- 10.0\,keV energy range. The errors were estimated using the {\tt cflux} and {\tt flux err} commands in {\tt xspec}. The relative errors on the fluxes are very similar to those on the source count rates.}
\end{table*}

Most O-type stars display a scaling relation between their X-ray and bolometric luminosity: $\log{\frac{L_{\rm X}}{L_{\rm bol}}} \simeq -6.5$ to $-7.2$ (Bergh\"ofer et al.\ \cite{Berghoefer}, Naz\'e \cite{Naze} and references therein). The relation is tight, but the exact value depends on the details of the analysis. To derive the $\log{\frac{L_{\rm X}}{L_{\rm bol}}}$ of the two Oe stars analysed here, we used the distance-independent ratio between X-ray and bolometric fluxes. The bolometric fluxes were obtained from the observed magnitudes and colours of the stars, and adopting the spectral types derived by Negueruela et al.\ (\cite{NSB}). For HD\,45314, we used the bolometric correction from Lanz \& Hubeny (\cite{LH2}) that corresponds to a temperature of 26000\,K and $\log{g} = 3.75$, whilst for HD\,60848 we adopted the bolometric correction from Lanz \& Hubeny (\cite{LH1}) for the same gravity, but a temperature of 31000\,K. In this way, we obtained $\log{\frac{L_{\rm X}}{L_{\rm bol}}} = -6.10 \pm 0.03$ for HD\,45314 and $-7.29 \pm 0.07$ for HD\,60848.

\section{X-ray variability}
We briefly consider the variability of the X-ray emission of the two Oe stars, on short and long timescales. 

\begin{figure}[h!tb]
\begin{center}
\resizebox{8cm}{!}{\includegraphics{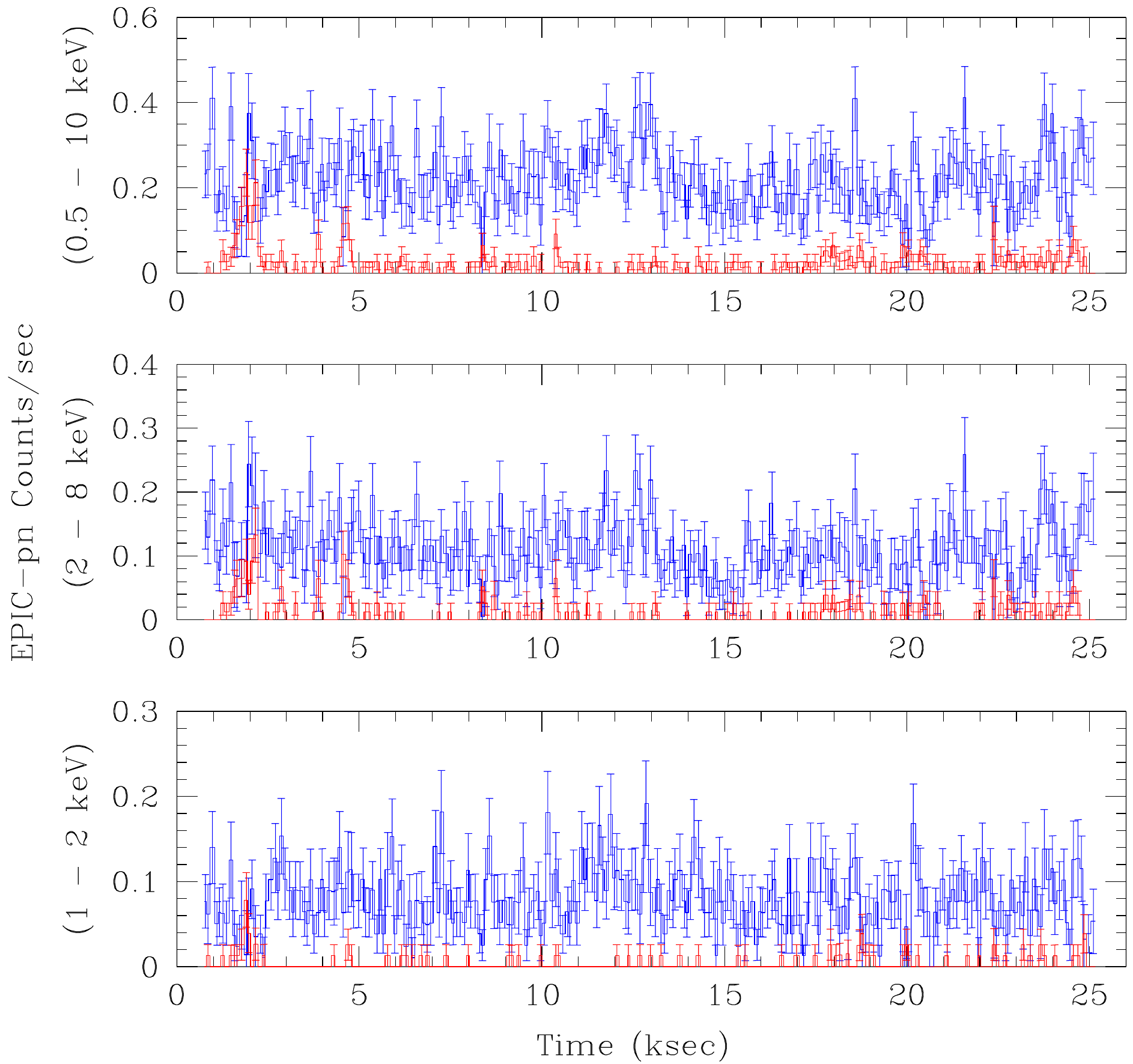}}
\end{center}
\caption{Background-subtracted EPIC-pn light curve of HD\,45314 with a bin of 100\,s (blue). The red histogram yields the light curve of the background.\label{fig2}}
\end{figure}
Regarding the short-term, intra-pointing variability, we restrict ourselves to HD\,45314, which has the higher count rate and the cleaner observation. Light curves with short time-bins ($ < 100$\,s) show variations, but they are similar to the error bars on the count rates, and are thus not significant. Figure\,\ref{fig2} illustrates the EPIC-pn light curve for time bins of 100\,s. As can be seen, the count rate varies by more than a factor 2 in a rather erratic way, although a $\chi^2$ test shows that these variations are significant at more than the 99.9\% level. A Fourier analysis of this light curve does not reveal a dominant frequency in these variations. This is again consistent with results found for $\gamma$\,Cas analogs (Smith et al.\ \cite{Smith1,Smith2}, Torrej\'on et al.\ \cite{TSN}).  

As for long-term variability, we have checked the HEASARC archives\footnote{http://heasarc.gsfc.nasa.gov/docs/archive.html} for previous observations of our targets. For HD\,45314, Chlebowski et al.\ (\cite{CHS}) quoted an upper limit of $12.5 \times 10^{-3}$\,cts\,s$^{-1}$ for the IPC instrument onboard {\it EINSTEIN}. The two Oe stars are apparent counterparts of {\it EXOSAT} sources within 6 and 10 - 15\arcsec for HD\,45314 and HD\,60848, respectively. The {\it EXOSAT} medium-energy proportional counter (ME) count rate is $0.13 \pm 0.05$ cts/s for HD\,45314. Unfortunately, the {\it EXOSAT} data of HD\,60848 are unreliable, because this observation was taken with the low-energy instrument and the 3000 Lexan filter, which was badly affected by UV leakage, especially for a bright, essentially unreddened O-star such as HD\,60848.

Adopting the parameters of our best-fit model for HD\,45314 and folding it through the IPC response matrix, we would expect a count rate of $14.5 \times 10^{-3}$\,cts\,s$^{-1}$, slightly above the upper limit of Chlebowski et al.\ (\cite{CHS}), but not inconsistent with the latter. We then used W3PIMMS to simulate an observation of HD\,45314 with the ME instrument onboard {\it EXOSAT}. This yields a predicted count rate of $7.3 \times 10^{-2}$\,cts\,s$^{-1}$, i.e.\ about half the observed one, although the latter has a large error-bar (see above). The X-ray emission of HD\,45314 seems thus relatively stable over long time scales. 

\section{Discussion and conclusions\label{conclusion}}
The X-ray observations presented here reveal that HD\,45314 is very likely a new member of the class of $\gamma$\,Cas analogs.

HD\,45314 might be a binary system. Mason et al.\ (\cite{Mason1}) reported speckle observations revealing a companion at 0.05\arcsec from HD\,45314, although the companion was not detected a decade later by Mason et al.\ (\cite{Mason2}), who argued that the companion could have moved closer to the main star. However, this companion is unlikely to be responsible for the $\gamma$\,Cas-like X-ray spectrum. Indeed, although Mason et al.\ (\cite{Mason1}) did not report an estimate of its magnitude, its detection probably rules out the possibility that it might be a white dwarf. Moreover, in such a wide orbit, accretion would certainly be very inefficient. 
In parallel, Boyajian et al.\ (\cite{Boyajian}) reported a jump in radial velocity between two observing runs separated by two months, which could indicate binarity with a period of a few months, although Boyajian et al. (\cite{Boyajian}) cautioned that the H$\alpha$ line profile on which they measured the radial velocity had changed between the two campaigns. Once again, the putative orbital period would probably be too long for accretion to be efficient.  

HD\,45314 most likely features a prominent decretion disk (Silaj et al.\ \cite{Silaj}), and this assumption is supported by the depolarization effect across the H$\alpha$ line reported by Vink et al.\ (\cite{Vink}). Therefore, the star-disk interaction scenario could explain the X-ray properties of HD\,45314. We stress that such an interaction is unlikely to take the form of a collision between magnetically channelled winds from the upper and lower stellar hemisphere with the dense equatorial disk, since strong magnetic fields and Keplerian Oe/Be disks are unlikely to coexist (ud-Doula et al.\ \cite{Asif}), because the fields would torque the material too much. Furthermore, to reach $kT = 21$\,keV by collisions of wind material with the disk would require a wind velocity of $\sim 4000$\,km\,s$^{-1}$, which seems rather unlikely. 

In $\gamma$\,Cas, the H$\alpha$ equivalent width correlates to first order with the variations of the $B-V$ colour index, and the column density towards the X-ray emission correlates with the $B-V$ index (Smith et al.\ \cite{Smith1}). Hence the X-ray absorption should correlate with the disk condition. For HD\,45314, we have found a significant but low column density in excess of the interstellar column. Still, it seems unlikely that this moderate column could totally hide lower-temperature plasma components, if they existed. Our optical data indicate that at the time of the X-ray observation, the disk had a relatively low density level, as judged from the strength of the H$\alpha$ emission, compared with the range of values seen by Rauw et al.\ (\cite{ibvs}). Hence, an even higher circumstellar column would be expected at times when the H$\alpha$ emission is stronger\footnote{Smith et al.\ (\cite{Smith1}) reported variations of the column density towards the harder X-ray component of $\gamma$\,Cas by at least one order of magnitude, whilst EW(H$\alpha$) changes by $\simeq 50$\%. Whether or not a similar relation holds for HD\,45314 is unclear, as we have only a single epoch of X-ray spectroscopy available.}. 

In conclusion, whilst the X-ray emission of HD\,60848 is quite typical for a late O-type star, HD\,45314 displays a very different behaviour, making it a new member of the class of $\gamma$\,Cas analogs. The most pressing question in this context is to understand the defining property that gives rise to a $\gamma$\,Cas-like behaviour. 
 
\acknowledgement{We thank the {\it XMM-Newton} SOC for their support, Andy Pollock for advice on the interpretation of the {\it EXOSAT} data, and Mira V\'eron (OHP) for accepting our ToO request for observations of HD\,45314. The Li\`ege team acknowledges support from the FRS/FNRS and from Belspo through XMM/INTEGRAL and GAIA-DPAC PRODEX contracts. AuD acknowledges support from NASA award NNX12AC72G.}


\begin{thebibliography}{}
\bibitem[1989]{AG} 
Anders, E., \& Grevesse, N.\ 1989, Geochimica et Cosmochimica Acta, 53, 197
\bibitem[1996]{Arnaud}
Arnaud, K.A.\ 1996, in Astronomical Data Analysis Software and Systems V, eds.\ Jacoby, G., \& Barnes, J., (San Francisco, ASP), 101, 17
\bibitem[1996]{Baranne} 
Baranne, A., Queloz, D., Mayor, M., et al.\ 1996, A\&AS, 119, 373
\bibitem[1997]{Berghoefer} 
Bergh\"ofer, T.W., Schmitt, J.H.M.M., Danner, R., \& Cassinelli, J.P.\ 1997, A\&A, 322, 167
\bibitem[2007]{Boyajian}
Boyajian, T.S., Gies, D.R., Baines, E.K., et al.\ 2007, PASP, 119, 742
\bibitem[1989]{CHS}
Chlebowski, T., Harnden, F.R.Jr., \& Sciortino, S.\ 1989, ApJ, 341, 427
\bibitem[1974]{CL}
Conti, P.S., \& Leep, E.M.\ 1974, ApJ, 193, 113 
\bibitem[2001]{RGS}
den Herder, J.W., Brinkman, A.C., Kahn, S.M., et al.\ 2001, A\&A, 365, L7
\bibitem[2012]{ISM}
Gudennavar, S.B., Bubbly, S.G., Preethi, K., \& Murthy, J.\ 2012, ApJS, 199, 8 
\bibitem[2009]{GN}
G\"udel, M., \& Naz\'e, Y.\ 2009, A\&ARv, 17, 309
\bibitem[2001]{Jansen}
Jansen, F., Lumb, D., Altieri, B., et al.\ 2001, A\&A, 365, L1
\bibitem[2003]{LH1} 
Lanz, T., \& Hubeny, I.\ 2003, ApJS, 146, 417
\bibitem[2007]{LH2} 
Lanz, T., \& Hubeny, I.\ 2007, ApJS, 169, 83
\bibitem[1998]{Mason1}
Mason, B.D., Gies, D.R., Hartkopf, W.I., Bagnuolo, W.G.Jr., Ten Brummelaar, T., \&  McAlister, H.A.\ 1998, AJ, 115, 821
\bibitem[2009]{Mason2}
Mason, B.D., Hartkopf, W.I., Gies, D.R., Henry, T.J., \& Helsel, J.W.\ 2009, AJ, 137, 3358
\bibitem[2009]{Naze}
Naz\'e, Y.\ 2009, A\&A, 506, 1055
\bibitem[2004]{wind}
Naz\'e, Y., Rauw, G., Vreux, J.-M., \& De Becker, M.\ 2004, A\&A, 417, 667
\bibitem[2010]{NRU}
Naz\'e, Y., Rauw, G., \& ud-Doula, A.\ 2010, A\&A, 510, 59
\bibitem[2011]{NRU2}
Naz\'e, Y., Rauw, G., \& ud-Doula, A.\ 2011, in Active OB stars: structure, evolution, mass loss and critical limits, Proc.\ IAU Symp.\ 272, eds.\ Neiner, C., Wade, G., Meynet, G., \& Peter, G., (Cambridge University Press, Cambridge), 624
\bibitem[2004]{NSB} 
Negueruela, I., Steele I.A., \& Bernabeu, G.\ 2004, AN, 325, 749
\bibitem[2008]{Sophie}
Perruchot, S., Kohler, D., Bouchy, F., et al.\ 2008, in Ground-based and Airborne Instrumentation for Astronomy II, eds. I.S.\ McLean, \& M.M.\ Casali, Proceedings of the SPIE, 7014, 70140J 
\bibitem[2003]{PR} 
Porter, J.M., \& Rivinius, T.\ 2003, PASP, 115, 1153
\bibitem[2006]{Rakowski}
Rakowski, C.E., Schulz, N.S., Wolk, S.J., \& Testa, P.\ 2006, ApJ, 649, L111
\bibitem[2007]{ibvs}
Rauw, G., Naz\'e, Y., Marique, P.X., et al.\ 2007, IBVS, 5773, 1
\bibitem[2010]{Silaj}
Silaj, J., Jones, C.E., Tycner, C., Sigut, T.A.A., \& Smith, A.D.\ 2010, ApJS, 187, 228
\bibitem[2012a]{Smith1}
Smith, M.A., Lopes de Oliveira, R., Motch, C., et al.\ 2012a, A\&A, 540, A53
\bibitem[2012b]{Smith2}
Smith, M.A., Lopes de Oliveira, R., \& Motch, C.\ 2012b, ApJ, 755, 64
\bibitem[2001]{apec} 
Smith, R.K., \& Brickhouse, N.S.\ 2001, ApJ, 556, L91
\bibitem[2001]{pn}
Str\"uder, L., Briel, U., Dennerl, K., et al.\ 2001, A\&A, 365, L18
\bibitem[2012]{TSN}
Torrej\'{o}n, J.M., Schulz, N.S., \& Nowak, M.A.\ 2012, ApJ, 750, 75
\bibitem[2001]{MOS}
Turner, M.J.L., Abbey, A., Arnaud, M., et al.\ 2001, A\&A, 365, L27
\bibitem[2006]{Asif}
ud-Doula, A., Townsend, R.H.D., \& Owocki, S.P.\ 2006, ApJ, 640, L191
\bibitem[2009]{Vink}
Vink, J.S., Davies, B., Harries, T.J., Oudmaijer, R.D., \& Walborn, N.R.\ 2009, A\&A, 505, 743
\end{thebibliography}
\end{document}